# Direct Low Temperature Nano-Graphene Synthesis over a Dielectric Insulator

*Mark H. Rümmeli*[*,†,‡], *Alicja Bachmatiuk*[†], *Andrew Scott*[†,§], *Felix. Börrnert*[†], *Jamie H. Warner*[¥], *Volker Hoffman*[†], *Jarrn-Horng Lin*[∫], *Gianaurelio Cuniberti*[§], *Bernd Büchner*[†],

IFW Dresden, P.O. Box 270116, 01171 Dresden, Germany

Department of Physics, Technische Universität Dresden, 01062 Dresden, Germany

Institute for Materials Science and Max Bergmann Center of Biomaterials Dresden University of Technology Dresden, 01062, Germany

Department of Materials, University of Oxford, Parks Rd, Oxford, OX1 3PH, United Kingdom

Department of Material Science, National University of Tainan, 33, Sec. 2, Shu-Lin Street, Tainan, Taiwan 700, Republic of China

Addresses correspondence to *m.ruemmeli@ifw-dresden.de

**ABSTRACT.** Graphene ranks highly as a possible material for future high-speed and flexible electronics. Current fabrication routes, which rely on metal substrates, require post synthesis transfer of the graphene onto a Si wafer or in the case of epitaxial growth on SiC, temperatures above 1000 °C are required. Both the handling difficulty and high temperatures are not best suited to present day silicon technology. We report a facile chemical vapor deposition approach in which nano-graphene and few layer graphene is directly formed over magnesium oxide and can be achieved at temperatures as low as 325 °C.

Graphene is a remarkable material with incredible electrical and mechanical properties. However, it has only recently been isolated.[1] This has made graphene the "new rising star" in nano-carbon based materials due to its exciting properties at the nanoscale, e.g. high charge carrier mobility.[2] In addition, when formed into narrow strips or ribbons (ca. 10 nm wide) a band gap opens making them excellent candidates for field effect transistors.[3] Hence, apart from the exciting possibilities in discovering new chemistry and physics from these 2D structures, they offer tantalizing opportunities for the development of high speed (and even flexible) molecular electronics. However, one of the major barriers impeding their progress on this front relates to difficulties in their fabrication. In order that graphene can make significant improvements to present day technologies, it needs to be synthesized in a manner suitable for planar fabrication technologies where reproducibility is essential. Randomly placed and structured graphene flakes are not suitable for this. Epitaxial grown graphene from SiC wafers seems a promising route for large area growth.[4] Chemical vapor deposition (CVD) grown graphene on metals has the drawback that the graphene needs to be transferred onto a wafer after synthesis.[5] A possible alternative route is the use of oxides to directly form graphene layers via CVD. This is attractive because most of today's transistor technology uses complementary metal-oxide semiconductor (CMOS) technology in which an oxide layer insulates the transistor gate from the channel. Hence, the ability to synthesize graphene directly on an oxide crucially removes the need to transfer the graphene after synthesis and can remove the need for large area synthesis as required with metal substrates.[6] Moreover, in order that the technique could easily be adapted for use in Si based technology low temperature reactions (400-450 °C) are required to maintain the mechanical integrity of low dielectric constant (K) intermetal dielectrics.[7] We have previously shown the potential for oxides to graphitize carbon.[8,9] Recent studies using silicon oxide[10,11] and zirconia[12] to grow single walled carbon nanotubes corroborate oxides can graphitize carbon. In this report we conclusively demonstrate MgO is suitable for the direct fabrication of graphene and few layer graphene (FLG) via CVD. Moreover, adjusting the reaction time or temperature allows one to switch between FLG and graphene formation. We have achieved synthesis temperatures down to 325 °C using acetylene as the feedstock.

**RESULTS AND DISCUSSION.**

Figure 1 presents a series of low voltage third order aberration corrected HRTEM data from samples prepared via catalytic CVD reactions on MgO using cyclohexane as the feedstock. Figure 1 a & b presents a MgO crystal after a CVD reaction in cyclohexane at 875 °C with a reaction time of 5 min. Graphitic layers (2-10) are seen on the surface of the nano-crystals and have an interspacing of *ca*. 3.5 Å. The graphitic layers align themselves with the MgO [100] lattice planes (spacing 0.21 nm), bending if necessary to do so. In other words, the graphene layers anchor to the MgO crystal. Most graphene layers anchor into every second lattice plane. However, every so often they anchor to a consecutive plane, presumably to minimize the strain. This is similar to the manner in which graphitic layers formed through SiC decomposition anchor to SiC.[13] For reaction times between 5 min. and 1 h the number of graphene layers formed is always the same, ranging between 2 and 10 layers. By reducing the reaction time one can tune the reaction to yield single graphene layers on the surface. Figures 1 c through f show cross sectional views and top views of graphene islands on the surface of the oxide crystals, respectively. To confirm the formation of graphene the samples were immersed in HCl (5M) to dissolve the MgO out. The process leaves only nano-graphite or nano-graphene shells as shown in figure 1 g and h.

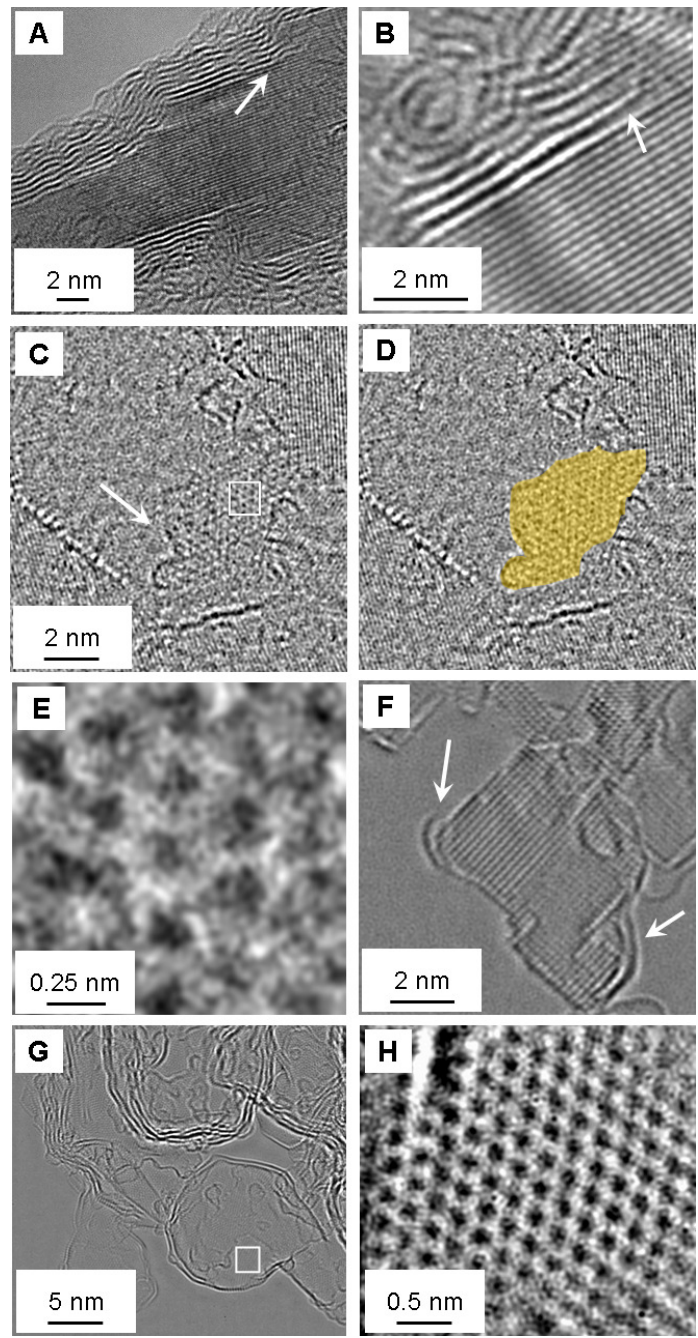

*Figure 1.* A. TEM image of few layer graphene on MgO crystal. Note the graphene layers interface directly to the MgO lattice fringes (A & B) [cyclohexane, 875 °C, 5 min]. C. & D. Graphene island on the surface of a MgO crystal. [cyclohexane, 875 °C, 30 s] E. Magnified region from box in panel C highlighting the graphene structure. F. Cross-section view of graphene on the surface of a MgO crystal. G. graphene shells after removal of MgO. [cyclohexane, 875 °C 1 h] H. Magnified region from box in panel C highlighting graphene structure.

The Raman spectra (Figure 2) from the samples show the G mode (due to bond stretching between pairs of $sp^2$ carbon atoms) and the D peak (due to breathing modes of $sp^2$ carbon atoms in rings), which lie around 1580 cm$^{-1}$ and 1360 cm$^{-1}$ respectively and provide further confirmation for graphitic carbon formation. Usually for graphene and few layer graphene one obtains an intense and narrow G band, a weak D mode and a very intense 2D mode at ca. 2700 cm$^{-1}$. The 2D mode is the second order D mode and its intensity is dependent on the number of layer. [Resp1] These features are clearly not what we obtain from our samples as can be seen in figure 2 a and b. Instead, we obtain a strong D mode and a broadened G mode and a weak and broad D mode. These differences occur due to the samples being comprised of nano-graphene. The nano-sized graphene domains have a large number of edge states (circumference) relative to the bulk graphene structure (see figure 2 d). Ferrari discusses in detail the effects of edge states which are essentially defects. [resp1] He states that the edge states give rise to a D band. In addition the G band is broadened and the average G position shifts up from around 1580 cm$^{-1}$ to 1600 cm$^{-1}$. In addition the doublet structure of the D and 2D mode is lost. These effects are exactly what we observe in our Raman spectra and are fully concomitant with nano-graphitic species. In addition, Cançado et al. have shown that one can successfully determine the crystallite size, $L_a$ (nm), of nano-graphite by Raman spectroscopy using the general equation $L_a = (2.4 \times 10^{-10})\lambda_l^4 (I_d/I_g)^{-1}$, where $\lambda_l$ is the laser energy in nm, Id and Ig are the intensity of the D and G modes, respectively. [Resp2]

From the above equation we obtain nano-graphene domain sizes of around 50 nm. This is reasonable given the MgO nanocrystals over which the nano-graphene and few layer nano-graphene are grown vary in size from a several nm to a few hundred nm.

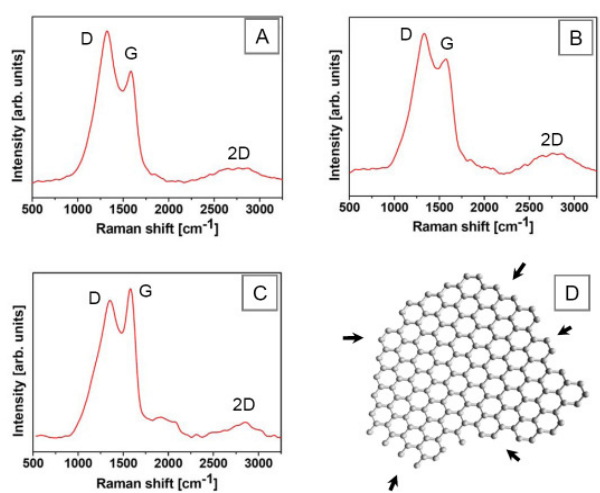

*Figure 2.* A - B. Raman spectra of nano-graphite from a purified sample prepared at 875 °C over MgO with a cyclohexane feedstock. Reaction time for A was 30 sec. and B was 5 min. C. Raman spectrum from purified sample prepared at 325 °C over MgO with cyclohexane as the feedstock and a reaction time of 10 min. D. Schematic of a nano-graphene flake illustrating the large number of edge defects relative to the bulk nano-graphene sheet. The high number of edge defects lead to a high D band.

The use of $C_2H_2$ for the low temperature CVD synthesis of carbon nanotubes has previously been shown down temperatures of 350 °C.[14] The ability to synthesize graphene via CVD at low temperature is important for direct device integration. Hence, we also explored the use of acetylene for graphene formation on MgO at 370 °C and 325 °C. With a reaction time of 10 min., we obtained nano-sized graphene islands on the surface. Some islands could be observed to align with the [100] lattice planes of the crystal. Figure 3 shows these traits for samples synthesized at 370 °C (panel a) and 325 °C (panel b). The data show the successful formation of graphene at 325 °C. To confirm the presence of graphene the samples were purified for further investigation. TEM studies showed the purified samples to consist of nano-graphite. Figure 3 c shows a typical sample in which various Moiré patterns can be observed. Fast Fourier transform data from the acquired micrographs consistently showed spot sequences and spacing arising from the [100] graphite lattice plane (e.g. figure 2 d). In addition, measurements showed interlayer spacing of 3.5 Å concomitant with the interlayer spacing in graphite as shown, for example, in panel e. Moreover, the Raman spectra (figure 2c) showed the presence of the G and D modes authenticating $sp^2$ carbon synthesis at the low temperature of 325 °C. The mean crystallite sizes, estimated from the Raman spectra as discussed above were around 30 nm. This is larger than the estimates from TEM for the nano-graphene flakes which range from 2 to 5 nm. The discrepancy may be due to nano-graphene islands coalescing together during the purification process. The coalescence of the nano-graphene islands reduces the number of dingling bonds (edge states) and hence reduces energy.

The effect of extended reaction times (up to 2 h) was also investigated. The results showed an increase in the number islands on the surface of the MgO nano-crystals. For samples formed with a 2 h reaction time, occasionally 2 or 3 graphitic layers could be observed. However, for the most part, the surface was covered with nano-graphene islands as illustrated in figure 2 f and g.

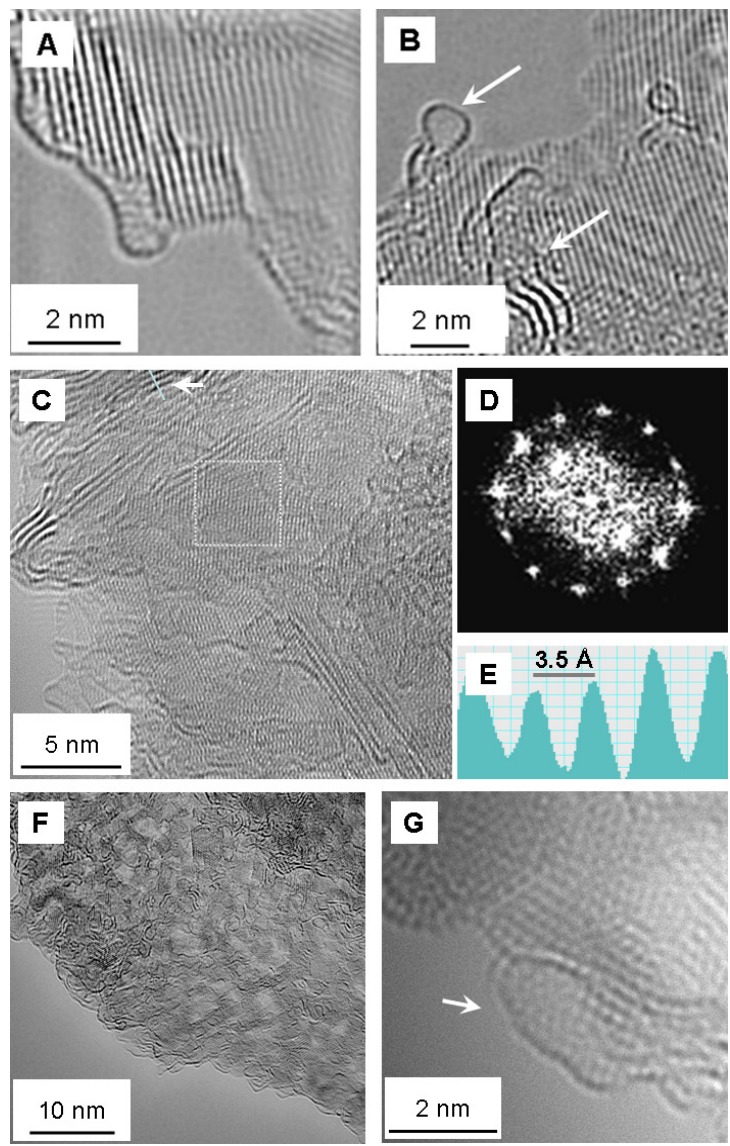

*Figure 3.* A. Graphene layer at the edge of a MgO crystallite. [acetylene, 370 °C, 10 min.] B. Graphene layer interfacing with MgO lattice fringes and graphene nano-islands on the surface/edge. [acetylene, 325 °C, 10 min.] Note: The images have been reconstructed using a mask applied to a 2D fast Fourier transform of the image. C. Purified graphite from graphene formed on MgO [acetylene, 325 °C, 10 min.]. D. Fast Fourier transform image from region (square) marked in panel D showing the diffraction patterns from (100) graphite. E. Contrast from interlayer graphite taken from panel C (see arrow). F. Dense nano-graphene island formation on MgO after a 2 h reaction period [acetylene, 325 °C, 2 h] G. Nano-graphene island on the edge of an MgO nano-crystal [acetylene, 325 °C, 2 h].

## CONCLUSION

The results clearly demonstrate that graphene can be grown directly on MgO. Similar to earlier FLG studies, samples formed at higher temperatures show graphene layers anchored to crystal planes.[9,13] Moreover, with the higher reaction temperature of 875 °C, the number of layers is always limited regardless of reaction time. This hints that the incorporation of carbon atoms to the graphene network occurs at the crystal/graphene layer interface, *viz.* bottom-up growth. Such a growth process is fundamentally different to the growth mechanisms argued for other graphene routes. As pointed out by Sutter,[4] the development of an atomically precise 'bottom-up' synthesis of graphene nanostructures would be a true leap forward. However, the presented data does not conclusively show this. The large number of graphene islands formed at lower temperatures with longer reaction times indicates island coalescence via Smoluchowski ripening cannot be ruled out.[15]

Our results show that MgO is suitable as a support for the synthesis of graphene via CVD. Moreover the route allows graphene to be formed at low temperatures down to 325 °C. The technique can avoid the need for post-synthesis transfer. Moreover, because it is also accessible at low temperatures it holds promise for the fabrication of large area, as well as nano-ribbon graphene using current Si based technologies.

## EXPERIMENTAL

The CVD setup consisted of a purpose built horizontal tube furnace with a quartz tube reactor. High purity MgO nano-crystal powder (Alfa Aesar, purity: 99.99%) was placed in an alumina crucible. Prior to synthesis the reactor was first evacuated down to 1 hPa, after which the hydrocarbon was introduced. Both static feedstock conditions (cyclohexane at 10 kPa, static flow) and flowing conditions (acetylene/argon at 100 kPa, 600 sccm,) were used. Synthesis temperatures between 875 °C and 325 °C were explored. Reaction times between 1 h and 10 s. were investigated. The as produced samples were investigated using 3rd order aberration corrected transmission electron microscopy (FEI Titan 300-80) operating at 80 kV. Raman spectroscopy was conducted with a ThermoScientific SmartRaman spectrometer. The available lasers are 780 nm, 633 nm, and 532 nm.

**Acknowledgment.** MHR thanks the EU and the Free State of Saxony for support via ECEMP. AS thanks the EU for support via its ERASMUS program. We are grateful to S. Leger, R. Schönfelder, M. Ulbrich and R. Hübel for technical support.

**Supporting Information Available:** Raman spectra. This material is available free of charge via the Internet at http://pubs.acs.org.

**References**

1. Novoselov, K. S.; Geim, A. K.; Morozov, S. V.; Jiang, D.; Zhang, Y.; Dubunos, S. V.; Grigoriava, I. V.; Firsov, A. A. Electric Field Effect in Atomically Thin Carbon Films. *Science* **2004**, 306, 666-669.

2. Novoselov, K. S.; Geim, A. K.; Morozov, S.; Jiang, D.; Katsnelson, M. I.; Grigorieva, I. V.; Dubonos, S. V.; Firsov, A. A. Two-dimensional gas of massless Dirac fermions in graphene. *Nature* **2005**, 438, 197-200.

3. Ponomarenko, L. A.; Schedin, F.; Katsnelson, M. I.; Yang, R.; Hill, E. W.; Novoselov, K. S.; Geim, A. K. Chaotic Dirac Billiard in Graphene Quantum Dots. *Science* **2008**, 320, 356-358.

4. Sutter, P. Epitaxial graphene: How silicon leaves the scene. *Nature Mater.* **2009**, 8, 171-172.

5. Obraztsov, A. N. Chemical vapour deposition: Making graphene on a large scale. *Nature Nano.* **2009**, 4, 212-213.

6. Li, X. S.; Cai, W. W.; An, J. H.; Kim, S.; Nah, J.; Yang, D. X.; Piner, R. D.; Velamakanni, A.; Jung, I.; Tutuc, E.; Banerjee, S. K.; Colombo, L.; Ruoff, R. S. Large-Area Synthesis of High-Quality and Uniform Graphene Films on Copper Foils. *Science* **2009**, 324, 1312-1314.

7. Morgen, M.; Ryan, E. T.; Zhao, J. H.; Hu, C.; Cho., T. H.; Ho, P. S. Low Dielectric Constant Materials for ULSI Interconnects. *Annu. Rev. Mater. Sci*. **2000**, 30, 645-680.

8. Rümmeli, M. H.; Borowiak-Palen, E.; Gemming, T.; Pichler, T.; Knupfer, M.; Kalbác, M.; Dunsch, L.; Jost, O.; Silva, S. R. P.; Pompe, W.; Büchner, B. Novel catalysts, room temperature, and the importance of oxygen for the synthesis of single-walled carbon nanotubes. *Nano. Lett.* **2005**, 5, 1209-1215.

9. Rümmeli, M. H.; Kramberger, C.; Grüneis, A.; Ayala, P.; Gemming, T.; Büchner, B.; Pichler, T. On the Graphitization Nature of Oxides for the Formation of Carbon Nanostructures. *Chem. Mater.* **2007**, 19, 4105-4107.

10. Liu, B.; Ren, W.; Gao, L.; Li, S.; Pei, S.; Liu, C.; Jiang, S.; Cheng, H. M. Metal-Catalyst-Free Growth of Single-Walled Carbon Nanotubes. *J. Am. Chem. Soc.* **2009**, 131, 2082-2083.

11. Huang, S.; Cai, Q.; Chen, J.; Qian, Y.; Zhang, L. Metal-Catalyst-Free Growth of Single-Walled Carbon Nanotubes on Substrates. *J. Am. Chem. Soc*. **2009**, 131, 2094-2095.

12. Steiner, III, S. A.; Baumann, T. F.; Bayer, B. C.; Blume, R.; Worsley, M. A.; MoberlyChan, W. J.; Shaw, E. L.; Schloegl, R.; Hart, A. J.; Hofmann, S.; Wardle, B. L. Nanoscale Zirconia as a Nonmetallic Catalyst for Graphitization of Carbon and Growth of Single- and Multiwall Carbon Nanotubes. *J. Am. Chem. Soc.* **2009**, 131, 12144-12154.

13. Kusunoki, M.; Rokkaku, M.; Suzuki, T. Epitaxial carbon nanotube film self-organized by sublimation decomposition of silicon carbide. *Appl. Phys. Lett.* **1997**, 71, 2620-2622.

14. Cantoro, M.; Hofmann, S.; Pisana, S.; Scardaci, V.; Parvez, A.; Ducati, C.; Ferrari, A. C;. Blackburn, A. A. M.; Wang, K. Y.; Robertson, J. Catalytic Chemical Vapor Deposition of Single-Wall Carbon Nanotubes at Low Temperatures. *Nano Lett.* **2006**, 6, 1107-1112.

15. Coraux, J.; N'Diaye, A. T.; Engler, M.; Busse, C.; Wall, D.; Buckanie, N.; Meyer zu Heringdorf, F-J.; van Gastel, R.; Poelsema, B.; Michely, T. Growth of graphene on Ir(111). *New J. Phys.* **2009**, 11, 023006/1-023006/22.

ToC graphic

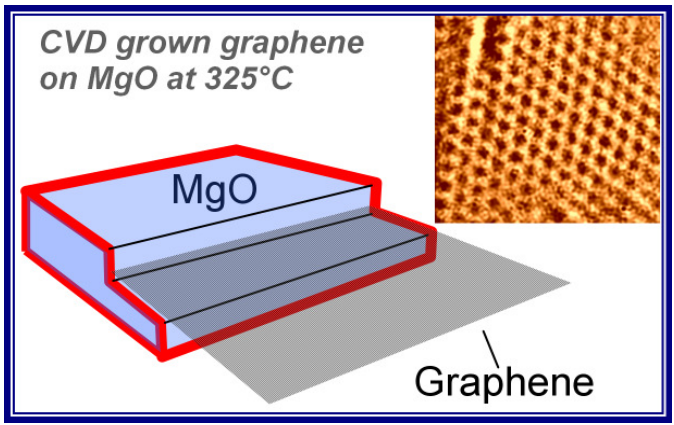

Graphene over MgO via CVD at 325°C is demonstrated. The direct synthesis of graphene on oxides may remove the need for large area graphene synthesis and its subsequent transfer on to silicon wafers.